# Constraints on the Origins of the Galactic Neutrino Flux Detected by IceCube

Abhishek Desai[1], Justin Vandenbroucke[1], Samalka Anandagoda[2], Jessie Thwaites[1], and M. J. Romfoe[1]
[1] Dept. of Physics and Wisconsin IceCube Particle Astrophysics Center, University of Wisconsin–Madison, Madison, WI 53706, USA; adesai@icecube.wisc.edu
[2] Clemson University, Department of Physics & Astronomy, Clemson, SC 29634-0978, USA

## Abstract

Galactic and extragalactic objects in the Universe are sources of high-energy neutrinos that may contribute to the astrophysical neutrino signal seen by IceCube. Recently, a study done using cascade-like events seen by IceCube reported neutrino emission from the Galactic plane with >4σ significance. In this work, we put a lower limit on the number of Galactic sources required to explain this emission. To achieve this, we use a simulation package created to simulate point sources in the Galaxy along with the neutrino and gamma-ray flux emissions originating from them. Along with using past IceCube discovery potential curves, we also account for Eddington bias effects due to Poisson fluctuations in the number of detected neutrino events. We present a toy Monte Carlo simulation to show that there must be at least eight sources, each with luminosity less than $1.6 \times 10^{35}$ erg s$^{-1}$, responsible for the Galactic neutrino emission. Our results constrain the number of individual point-like emission regions, which apply both to discrete astrophysical sources and to individual points of diffuse emission.

## 1. Introduction

The Milky Way is host to a variety of astrophysical objects, interstellar gas, and radiation fields. By observing the particles created through interactions within and between these phenomena, we can deepen our understanding of the processes involved. While photons from the Milky Way are easily observable—making the Galactic plane the brightest region in the sky—other particles like neutrinos are not so easily observable. Neutrinos can be produced via processes like stellar explosions or supernovae (see, e.g., Thompson et al. 2003), the interaction of cosmic rays with matter (Domokos et al. 1993), binary systems with a compact object and a massive star (Levinson & Waxman 2001; Kheirandish 2020), or other sources in our Galaxy. In this work, we study neutrinos in the TeV–PeV regime, which can be produced due to cosmic-ray interactions or sources like pulsars and supernova remnants in the Galaxy.

High-energy neutrinos are produced from the decay of charged pions ($\pi^{\pm}$), which are the result of hadronic ($pp$) or photohadronic ($p\gamma$) interactions. These processes also lead to the production of neutral pions ($\pi^0$), which then decay into gamma rays. The Galactic plane dominates the observed gamma-ray sky and is studied in the GeV to TeV regime by various observatories like Fermi-LAT (Ackermann et al. 2012, 2015), HAWC (Zhou et al. 2017; Albert et al. 2020), HESS (Abdalla et al. 2018), TIBET (Amenomori et al. 2021), LHAASO (Cao et al. 2023), MAGIC (Acciari et al. 2020), VERITAS (Adams et al. 2021), etc. As only hadronic interactions can give rise to both neutrinos and gamma rays, a multimessenger detection using neutrino and gamma-ray observations will allow us to investigate the fundamental processes behind these interactions in the Galactic plane.

The IceCube neutrino observatory is an in-ice cubic kilometer neutrino detector at the South Pole, which detects high-energy neutrinos through their interactions in the ice (Aartsen et al. 2017b). Muon neutrino charged-current interactions give rise to muons that are long lived and travel several kilometers in the ice. On the other hand, short-lived hadronic cascades in the ice can be created through all-flavor neutral current interactions or through electron and tau neutrino charged-current interactions. These interactions in the ice emit Cherenkov radiation that is detected by digital optical modules along strings embedded in the ice. The IceCube data acquisition system estimates the detected event energy, direction, and event morphology, which is track-like or cascade-like (Abbasi et al. 2009; Aartsen et al. 2017b).

Identification of the production mechanisms and sources of high-energy neutrinos observed by IceCube is one of the prime questions of multimessenger astrophysics. Recently, Abbasi et al. (2023a) referred to as deep neural network (DNN) cascades from this point) reported neutrino emission with >4σ significance from the Galactic plane by making use of cascade-like events detected by the IceCube Neutrino Observatory. Detection of this Galactic neutrino signal raises important questions analogous to those posed by the isotropic (extragalactic) neutrino signal.

Although there is evidence for two individual extragalactic neutrino sources, TXS 0506+056 (Aartsen et al. 2018) and NGC 1068 (Abbasi et al. 2022a), they only contribute ∼1% (depending on energy) of the total extragalactic signal (Abbasi et al. 2022a). Nondetection of a larger number of extragalactic sources constrains the emission to be produced by an abundant population of relatively low-luminosity sources (Abbasi et al. 2023b). However, both the number density and luminosity function of the sources as a whole remain unknown, along with the extent to which these quantities evolve with redshift and what the distance scale of typical sources is. For the Galactic neutrino signal, on the other hand, even before determining the nature and origins of the emission, the distance scale and even the approximate large-scale spatial distribution of the emission

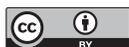

**Table 1**
Number of Sources ($N_{src}$) Making up the Measured Best-fit Total Galactic Neutrino Flux (in units of TeV$^{-1}$ cm$^{-2}$ s$^{-1}$) Assuming All Sources Are Point Sources Located at the Center of the Galaxy (at $\sim-28°$ decl.) and Have Flux Values Given by the Sensitivity and Discovery Potential Curves from the DNN Cascades and PS Tracks Work

| Sample Tested | $E^{-2.0}$ Flux at $\sim-28°$ | $>N_{src}$ ($E^{-2.0}$) | $E^{-3.0}$ Flux at $\sim-28°$ | $>N_{src}$ ($E^{-3.0}$) |
| --- | --- | --- | --- | --- |
| DNN 4$\sigma$ DP | $3.26 \times 10^{-16}$ | 5 | $1.97 \times 10^{-16}$ | 8 |
| PST 5$\sigma$ DP | $7.97 \times 10^{-16}$ | 2 | $1.37 \times 10^{-14}$ | 0 |

**Note.** Discovery potential is denoted DP, DNN cascades is denoted DNN, and PS tracks is denoted PST. The flux spectrum ($dN/dE$) proportional to $E^{-2}$ and $E^{-3}$ is tested for both cases. As no Galactic neutrino sources have been detected, these can be viewed as lower limits on the number of such sources. A more detailed study is described in Section 5 to understand these numbers better. We also report in Table 3 in the Appendix how $N_{src}$ changed when sensitivity curves are used instead of DP.

is well known. This means that the IceCube measurement of the total Galactic neutrino flux is also a measurement of the total Galactic neutrino luminosity. It can also be used to put constraints on the typical luminosity of contributing sources which are directly related not only to the number density of sources but also to the total number of sources in the Galaxy.

Diffuse Galactic neutrino emission is generically expected from cosmic rays interacting with the interstellar medium to produce pions that decay to gamma rays and neutrinos. There are likely also individual astrophysical sources of neutrinos within the Galaxy (Kheirandish 2020). Searches for possible Galactic neutrino emitters using various candidate source lists have been performed, some examples of which include multiple Galactic source lists (Aartsen et al. 2017a), pulsar wind nebulae (PWNe; Aartsen et al. 2020a), HAWC Galactic sources (Kheirandish & Wood 2019), X-ray binaries (Abbasi et al. 2022b), and LHASSO sources (Abbasi et al. 2023c). No significant emission has been found, and all these studies placed upper limits on the neutrino flux from Galactic sources. However, the total Galactic flux reported in Abbasi et al. (2023a) is several times higher than predicted by models of diffuse emission, indicating that discrete sources may provide an important component of the flux. For particular Galactic source classes tested in Abbasi et al. (2023a), the neutrino source contributions cannot be distinguished from one source class to another, nor from the diffuse scenario. This is because the IceCube cascade sample has relatively large (∼7°) angular resolution and the large-scale spatial distribution of the emission is very similar in the various scenarios. Because of the spatial similarity, constraints on the luminosity and number of Galactic sources can be constructed robustly, with a weak dependence on the exact spatial distribution. In this work, we present a simulation package to estimate the neutrino (and gamma-ray) contribution from individual Galactic sources and use it to draw conclusions from the results presented by Abbasi et al. (2023a).

This work is divided as follows. Section 2 describes a simple limit on the number of sources contributing to the neutrino signal without the use of any simulation. Section 3 explains the simulation package and how it works. In Section 4, we discuss how the package can be used to simulate gamma-ray source populations, comparing them to detected sources. In Section 5, we use the simulation to determine a lower limit on the number of sources or emission sites for Galactic neutrinos. This is done by estimating the number of Galactic sources or neutrino emission points that can be detected by IceCube, the conclusions for which are explained in Section 6.

## 2. Neutrinos from the Galactic Center

We start with a simple assumption that all neutrino sources responsible for the DNN cascades detection are collected at the center of the Galaxy and are point sources. The KRA$_\gamma^{50}$ (Gaggero et al. 2015) best-fit flux reported by DNN cascades can then be approximated as the total neutrino flux from the Galactic center. The best-fit flux for the KRA$_\gamma^{50}$ template at 100 TeV is given by $\sim 1.5 \times 10^{-15}$ TeV$^{-1}$ cm$^{-2}$ s$^{-1}$ (Figure 5 of Abbasi et al. 2023a). Note that we use "flux" in this work to denote the differential neutrino number flux at 100 TeV in units of TeV$^{-1}$ cm$^{-2}$ s$^{-1}$ unless specified otherwise.

Next, we use the $4\sigma$ discovery potential curves for the DNN cascade sample and $5\sigma$ discovery potential curves for the 10 yr point-source tracks sample (Aartsen et al. 2020b), referred to as PS tracks from this point. The DNN cascade sample is made up of cascade-like events collected over 10 yr and reconstructed using a DNN (for more details regarding event properties see Abbasi et al. 2023a) while the PS tracks sample is made up of track-live events also collected over 10 yr (for more details regarding event properties see Aartsen et al. 2020b). The $4(5)\sigma$ discovery potential shows the flux required to detect the source at $4(5)\sigma$. Assuming all the neutrino sources are close to or at the Galactic center (at $\sim-28°$), we obtain the sensitivity and discovery potential flux at that decl. This flux is then compared with the total measured flux $\sim 1.5 \times 10^{-15}$ TeV$^{-1}$ cm$^{-2}$ s$^{-1}$ to derive the number of sources contributing to the signal, as shown in Table 1. We also show, in the Appendix, how the results change when the sensitivity estimates are used instead of the discovery potential. The sensitivity gives the 90% confidence level (CL) upper limit when the test statistic (TS) equals 0.

The number of neutrino point sources calculated from this test uses the assumption that all sources are close to each other and at $\sim-28°$ decl. We can already see from this simple test that because of the improved sensitivity of the DNN cascades sample in the Southern Hemisphere, it has a better chance of detecting neutrinos from the Galactic center. However, because the localization is better for track events (less than 1° for $\sim$100 TeV events; see Abbasi et al. 2021a) compared to cascade events ($\sim$7° at 100 TeV; see Abbasi et al. 2023a), it would be difficult to resolve the sources, even if they are detected. In other words, source confusion would be a serious challenge if the sources are tightly clustered at the Galactic center. On the other hand, if they are sufficiently clustered, then they could be detected as an aggregate excess (perhaps spatially extended) above the surrounding diffuse emission.

A more robust way to perform this test is to simulate sources in our Galaxy and use the DNN cascades and PS tracks results to see if any of the samples have a chance of detecting them. We perform this study by using the simulation code described below.

## 3. SNuGGY

The simulation package created to simulate point-like Galactic sources is named "Simulation of the Neutrino and Gamma-ray Galactic Yield" (SNuGGY) and is made available in Desai et al. (2023). This package is analogous to an existing simulation tool called FIRESONG (Tung et al. 2021), which simulates extragalactic gamma-ray and neutrino sources. The main functioning of the code can be divided into two parts, explained in the subsections below.

### 3.1. Simulating Source Positions

The core logic to simulate the Galactic source population is to use the 2D probability distribution function (PDF) of the number density of Galactic sources in Galactocentric coordinates. This can either be a simple 2D exponential PDF (identified as "exponential" below) or a modified 2D exponential PDF mimicking a more realistic distribution (identified as "modified_exponential" below). If the location of the source with respect to the center of the Galaxy is given by $R$ and the vertical height above the Galactic plane is given by $z$, the two number densities mentioned above are given by the following equations.

1. For the exponential setup, the 2D PDF is given by

$$\rho(R, z) = \rho_0 \, e^{-\frac{R}{R_0}} \, e^{-\frac{|z|}{z_0}}, \qquad (1)$$

where $\rho_0$ is the normalization parameter and $R_0$ and $z_0$ are the scale length and scale height, respectively. For the example here, we use the same parameters as the ones used by Winter et al. (2016) for the millisecond pulsar (MSP) distribution and are given by $R_0 = 3$ kpc and $z_0 = 0.6$ kpc. Note that while determining the $R$ and $z$ PDFs, the Jacobian is included. While this feature is present in the SNuGGY framework, we do not use the exponential setup in the study presented in Section 5 below, but add it here for completeness.

2. For the modified_exponential setup, the 2D PDF (also shown in Figure 1) is given by

$$\rho(R, z) = \rho_0 \left(\frac{R}{R_\odot}\right)^\alpha \exp\left(-\beta \frac{R - R_0}{R_\odot}\right) \exp\left(-\frac{|z|}{h}\right), \qquad (2)$$

where $\alpha$, $\beta$, and $h$ are parameters of the distribution. This equation is taken from Ahlers et al. (2016) where the ($\alpha$, $\beta$, $h$) parameters for the pulsar distribution (2, 3.53, 0.181) shown by Lorimer et al. (2006) and supernova remnant distribution (1.93, 5.06, 0.181) by Case & Bhattacharya (1998) are used as reference. Figure 1 shows the 2D histogram using the parameters for the pulsar distribution case. Note that the vertical distance ($r$) peaks at a value away from the center of the Galaxy as expected from the Jacobian factor and agrees with pulsar distribution studies like Yusifov & Kucuk (2004).

These PDFs are converted to inverse cumulative distribution functions, which are then used to sample a given number of sources (the PDF sampling method used is similar to Tung et al. 2021). For now, only the two abovementioned setups exist in the framework, with the possibility of adding more in the future if required. Both existing frameworks simulate a disk galaxy with the galactic bulge and thickness depending on the specified input parameters. Figure 1 (bottom) shows the simulated sources for the abovementioned simulation in International Celestial Reference System coordinates.

### 3.2. Deriving Flux Estimates

Next, we simulate the observed neutrino flux for every source. This is done by assigning the integrated neutrino luminosity (over 10 TeV–10 PeV, and in units of erg s$^{-1}$) of the source and using the distance and spectrum to derive the flux, assuming a simple power-law spectral model. Note that we use the word "luminosity" from this point to denote integrated luminosity over 10 TeV–10 PeV unless specified otherwise. To get the luminosity, either a standard candle (SC) approach is used where all sources have equal luminosity, or a log-normal (LN) distribution is used and described below.

1. *Standard candle luminosity*. For a set of $N$ simulated sources, we assign the simulated distance, $d_i$, for each source, $i$. Let the total Galactic flux be given by $\phi_{\text{Galactic}}$ and $L_{\text{SC}}$ be the SC luminosity per source. Assuming that the luminosity of each source is the same and sources are centered on the Galactic center (~8 kpc; Leung et al. 2022), the SC luminosity is given by

$$L_{\text{SC}} = \frac{\phi_{\text{Galactic}}}{\frac{N}{4\pi(8\text{kpc})^2}}. \qquad (3)$$

This method ensures that the SC luminosity is selected such that the sum of fluxes per source, derived using $L_{\text{SC}}$ and $d_i$, is close to the total Galactic flux $\phi_{\text{Galactic}}$. To derive the differential flux from the integrated luminosity, a power-law spectrum with an index $\gamma$ is used over an energy range given by $E_{\min}$ and $E_{\max}$, where all these parameters are used as inputs to the simulation. This method includes cosmic variance, which varies the total Galactic flux for each simulation while keeping the luminosity per source fixed and is called the "Standard-Candle" approach in the simulation code. Forced SC luminosity (special case of SC): we also give a special case of simulating SC sources using the "Forced_standardCandle" mode in the code. For this forced case, the sum of simulated fluxes is exactly equal to the Galactic diffuse flux $\phi_{\text{Galactic}}$. $L_{\text{SC}}$, in this case, is simulated by

$$L_{\text{SC}} = \frac{\phi_{\text{Galactic}}}{\sum_{i=1}^{N} \frac{1}{4\pi d_i^2}}. \qquad (4)$$

Note that this method is similar to the previous approach, but the actual simulated distance per source is used in the calculation instead of a fixed value of 8 kpc. Due to this difference, the previous method will cause the total simulated flux to be close to but not equal to the input $\phi_{\text{Galactic}}$.

2. *Log-normal luminosity*. This method is the most realistic setup to simulate luminosities. It uses a PDF similar to the one described by Ploeg et al. (2017). For a mean luminosity $L_0$ in units of erg s$^{-1}$, the PDF can be given by

$$P_{\text{LN}}(L) = \frac{1}{\sigma_L L \sqrt{2\pi}} \exp\left(\frac{-(\ln L - \ln L_0)^2}{2\sigma_L^2}\right), \qquad (5)$$

where the $\sigma_L$ parameter controls the width of the distribution. A very low $\sigma_L$ value will result in a simulated distribution where all the luminosities equal the mean luminosity, i.e., a SC approach (see Figure 2). In the event that a mean luminosity value is not specified

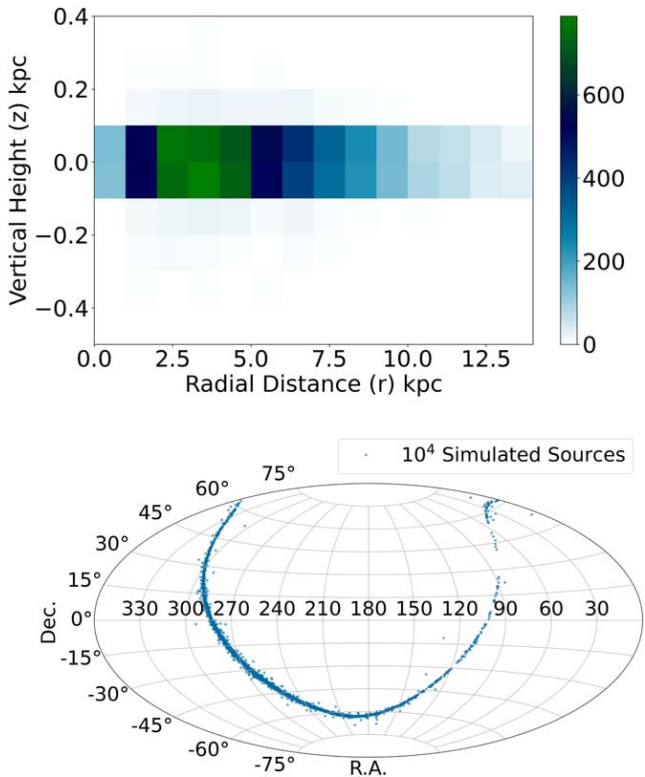

**Figure 1.** Top: 2D histogram showing the simulated vertical height (z) and distance (R) with respect to the center of the Galaxy for a set of $10^4$ sources. Bottom: simulated source positions in equatorial coordinates. Note that this is the result of one simulation while using the modified exponential distribution function described in Section 3.1, along with the pulsar distribution parameters.

as an input, the code uses the forced SC approach (Equation (4)) to first find the mean luminosity.

While the forced SC simulation ensures that the total simulated flux is exactly equal to the input total flux, the SC and LN cases vary the total simulated flux for every simulation. In a more realistic scenario, the total measured diffuse flux measurement is subject to cosmic variance due to uncertainties in the measurement. The SC and, more preferably, the LN simulation, accounts for this and should be used for a more nonbiased simulation study.

A comparison of the flux estimated using all three methods for one simulation of $10^4$ sources is shown in Figure 3. The total Galactic flux derived is also shown in the legend and matches exactly for the forced SC case. Note that for LN cases, the mean luminosity ($L_0$) is set to "None" so that the code uses an approach similar to Equation (4) to estimate $L_0$. A low value of $\sigma_L$ like 0.01 (as shown in the figure) will thus simulate a source population mimicking a SC approach. For the LN with higher $\sigma_L$, a more realistic scenario of the sum of fluxes not equaling the diffuse flux is seen.

These flux estimation methods can be used for simulating both neutrino and gamma-ray fluxes for sources. While features in the SNuGGY code have been added to simulate gamma-ray fluxes from neutrino fluxes (and vice versa) using $pp/p\gamma$ interactions (Kelner et al. 2006; Halzen 2022), neutrino and gamma-ray fluxes can also be simulated separately. This avoids relying on a model-dependent scenario where all neutrinos and gamma rays are related by $pp/p\gamma$ interactions. In this work we simulate the gamma-ray and neutrino fluxes individually.

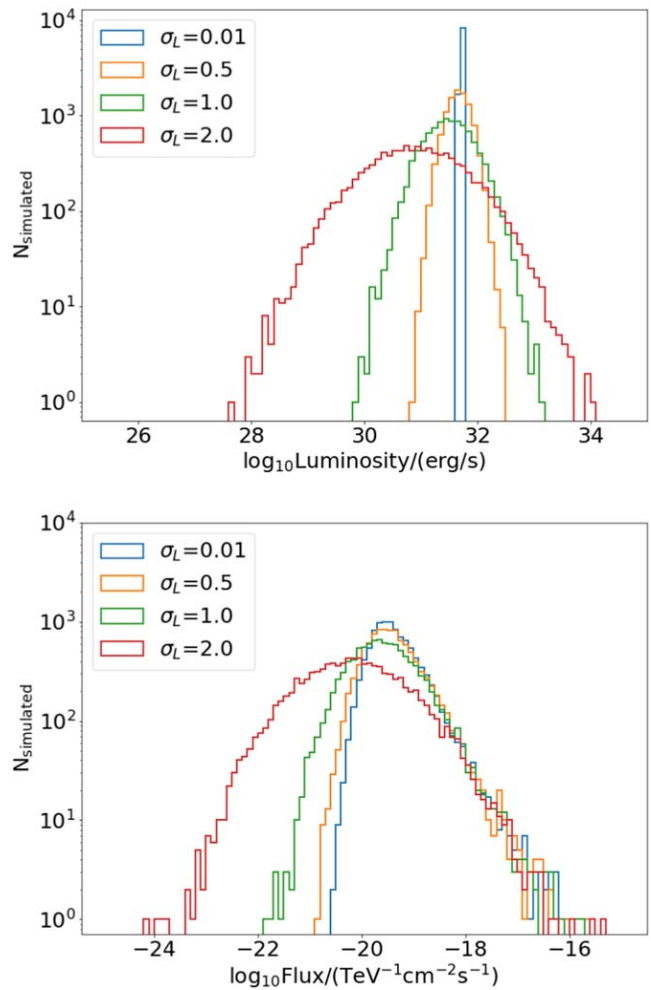

**Figure 2.** Top: luminosity distribution for sources simulated using the LN luminosity method. The mean luminosity in this example is calculated using Equation (4) where the total flux is fixed to $2.18 \times 10^{-15}$ TeV$^{-1}$ cm$^{-2}$ s$^{-1}$. For a low value of $\sigma_L = 0.01$, the distribution behaves like a SC where all source luminosities are close to the mean. Bottom: corresponding flux distribution for the simulated sources using the luminosities shown in the top plot. The flux values are the energy differential at 100 TeV. The flux calculation also uses an index of 2.0, an energy range of 0.1–100 TeV, and the source positions shown in Figure 1.

### 4. Checking SNuGGY Using Galactic Gamma-Ray Sources

The SNuGGY package can be used to simulate gamma-ray fluxes using the methods shown in Section 3.2. To test the validity of the code setup, we use the LN distribution method and compare it with existing gamma-ray observations of pulsar and pulsar-associated sources. While the source position distributions we use for the simulation follow a pattern similar to PWN sources, we are actually just simulating point sources using SNuGGY. This would mean we can compare observations of other pulsar source classes like MSPs provided that the source positions and luminosities are accounted for in the simulations.

*Test 1.* Check how the simulation behaves when the number of simulated sources is varied with the condition that the total simulated gamma-ray flux lies close to an input total gamma-ray flux. As the mean luminosity is not specified, the calculated mean luminosity (using Equation (4)) decreases as the number of sources increases. The total Galactic diffuse gamma-ray flux is estimated at 1 GeV using the model taken from Ackermann et al. (2015)

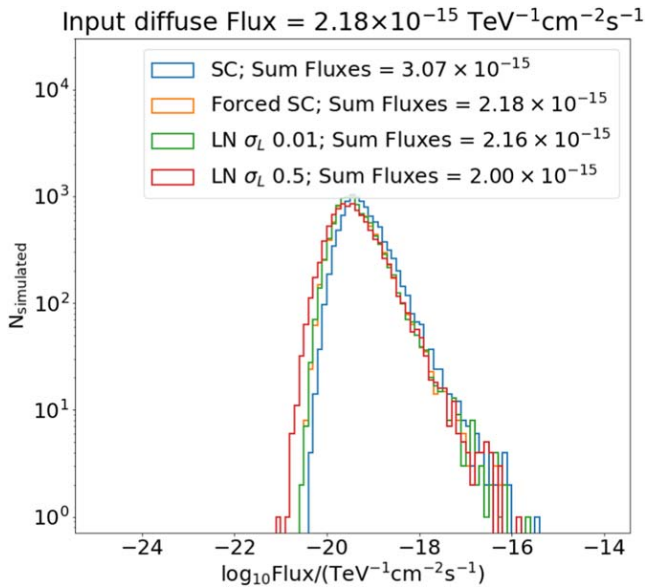

**Figure 3.** Flux distribution for $10^4$ sources simulated using different estimation techniques. The total sum of the simulated fluxes is also shown in the legend to compare to the input given for the total Galactic flux measurement of $2.18 \times 10^{-15}$ TeV$^{-1}$ cm$^{-2}$ s$^{-1}$ (Abbasi et al. 2023a). Note that the plot above depicts the results of one simulation and shows a simulation different from Figures 2, 4, and 5. This simulation is repeated 100 times to derive the mean total simulated flux for each of the cases, as follows: SC = $2.65 \times 10^{-15}$ TeV$^{-1}$ cm$^{-2}$ s$^{-1}$; forced SC = $2.18 \times 10^{-15}$ TeV$^{-1}$ cm$^{-2}$ s$^{-1}$; LN for $\sigma_L = 0.01$ of $2.18 \times 10^{-15}$ TeV$^{-1}$ cm$^{-2}$ s$^{-1}$; and LN for $\sigma_L = 0.5$ of $2.16 \times 10^{-15}$ TeV$^{-1}$ cm$^{-2}$ s$^{-1}$. After including uncertainties, these simulated fluxes match the measured total flux, given as $2.18^{0.53}_{0.49} \times 10^{-15}$ TeV$^{-1}$ cm$^{-2}$ s$^{-1}$ as reported by Abbasi et al. (2023a).

and given by $10^{-9}$ MeV$^{-1}$ cm$^{-2}$ s$^{-1}$. Using a $\sigma_L$ value of 0.01, we simulate a SC distribution of different numbers of sources (points of gamma-ray emission). The flux distribution of these simulations is then compared to the observations reported in the Fermi-LAT 4FGL-DR3 catalog (Abdollahi et al. 2022) for PWN and MSP sources. The observed sources fall on the bight end of the simulated source distribution (see Figure 4), highlighting the potential of using the SNuGGY framework for similar source population studies. Note that, for this sample case, we do not simulate a specific source class of pulsars but just a distribution of gamma-ray points in the Milky Way.

*Test 2*. Check how the simulation behaves when the luminosity function is changed with a fixed median luminosity. As the 3HWC catalog (Albert et al. 2020) reports observations for HAWC sources with corresponding TeV halo candidate pulsars within 1°of the observations, we use the reported luminosity estimates for this test. The high-energy ($\sim$7 TeV) observations of 11 pulsar sources seen by the HAWC telescope are used here. For this case, we use the mean luminosity $\sim 4.6 \times 10^{33}$ erg s$^{-1}$ (integrated over the energy range of 0.1–100 TeV) derived from the 3HWC catalog by using the reported flux measurements and distances given in Tables 2 and 4 of Albert et al. (2020). The simulated $10^4$ source flux distributions for a $\sigma_L$ value of 0.01 and 1.0 as compared to the reported HAWC observations at 7 TeV are shown in Figure 5. The estimated luminosities of the 3HWC sources vary from $\sim 1 \times 10^{31}$ to $\sim 1 \times 10^{35}$ erg s$^{-1}$ while the mean luminosity is $\sim 4.6 \times 10^{33}$ erg s$^{-1}$. This would mean that the luminosities follow a distribution similar to a LN distribution with a high $\sigma_L$ value. Using this in the simulation brings the simulation closer to the real scenario where the 3HWC sources should lie on the bright end of the histogram (see Figure 5).

We choose to use the Fermi-LAT measurements for Test 1 because of a better measurement of the total diffuse background as compared to the HAWC catalog. Conversely, we use the HAWC measurements for Test 2 because of the luminosity values reported in Albert et al. (2020; which span from $\sim 1 \times 10^{31}$ to $\sim 1 \times 10^{35}$ erg s$^{-1}$), which allows us to test a sample with a wider luminosity distribution. The luminosity function for low values of $\sigma_L$ does not impact the distribution greatly, but it starts to show an effect as $\sigma_L$ is increased (see Figures 3 and 5). Additionally, as the number of simulated sources increases, the brighter sources dominate the entire population (see Figure 4). We also see that the simulated population using SNuGGY is able to mimic the properties of the observed HAWC and Fermi-LAT samples for both tests depending on the input parameters used.

## 5. Galactic Neutrino Sources Analysis

We use the SNuGGY framework to simulate these Galactic neutrino sources of varying number densities and luminosities to understand better the Abbasi et al. (2023a) result. Abbasi et al. (2023a) report neutrino emission with >4σ significance from the Galactic plane but with no significant detection of a single Galactic source or subclass of Galactic sources.

All reported luminosities for the neutrino studies in this section are integrated luminosities over 10 TeV–10 PeV, while the fluxes (simulated and taken from Abbasi et al. 2023a) are differential measurements at 100 TeV. We test two cases for the source position: all simulated sources at the center and simulated sources following a PWN distribution simulated using $(\alpha, \beta, h) = (2, 3.53, 0.181)$ (Lorimer et al. 2006). Note that the spatial distribution of other Galactic neutrino source classes is similar to the PWN distribution, which allows us to make conclusions about Galactic neutrino source classes as a whole using this study. We derive the neutrino fluxes of these sources based on the SNuGGY framework. Once the sources are simulated with a corresponding neutrino flux, they are compared to the sensitivity and discovery potential curves from the IceCube samples (DNN cascades and PS tracks) to check whether the simulated flux is above the threshold for a source to be detected by IceCube. A source is considered to be detected if the simulated flux value is higher than the sensitivity or discovery potential value. This is shown as an example in Figure 6, where one simulation of $10^4$ neutrino sources and their fluxes are computed and shown as the blue data points and compared to the IceCube sensitivity curves. If the simulated flux is above the sensitivity curve, it shows that the neutrino sample is sensitive to neutrinos coming from that simulated source and will result in detection with a TS greater than the median TS. However, a comparison with the 4σ (5σ) discovery potential curves (instead of sensitivity) will allow us to find simulated sources that can be detected with a significance that is higher than 4σ (5σ).

### 5.1. Eddington Bias

We also account for Eddington bias while estimating the number of detected sources by accounting for Poisson fluctuations in the number of neutrinos per source. As described by Strotjohann et al. (2019), Eddington bias is the bias seen when upward Poisson fluctuations in the number of neutrinos detected from a source cause an overestimation of the source flux. This bias can be particularly large when there are many dim sources. Using the effective area curves reported by Abbasi et al. (2021a; IC86v2; for PS tracks) and by Abbasi et al. (2023a; their Figure 2; for DNN cascades), we estimate the number of expected neutrino events, over a period of 10 yr, depending on the simulated source flux. Note that for this calculation, we make the approximation that the effective area curves of the DNN cascade sample do not change with decl. While the number of events would change as the decl. changes, especially at higher energies, the change is not significant in our analysis and would only modify the uncertainty in the cases of a large number of simulated sources due to more sources being simulated across the Galactic plane and not just close to the center.

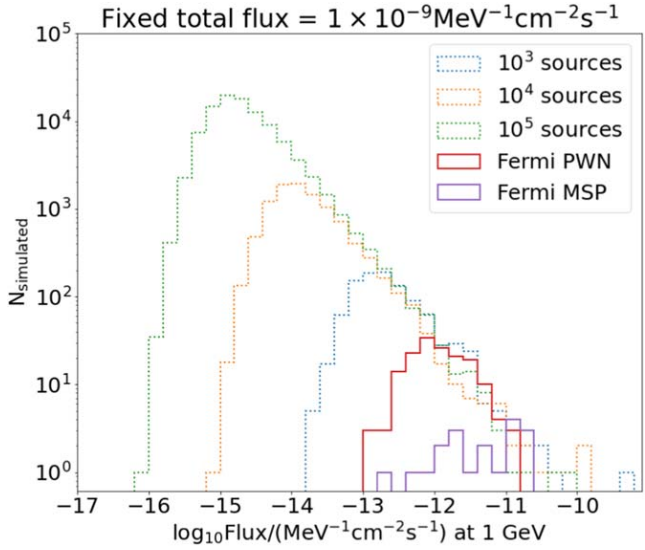

**Figure 4.** Point sources simulated using the Fermi-LAT Galactic diffuse flux measurement at 1 GeV in units of $MeV^{-1}$ $cm^{-2}$ $s^{-1}$ (only for this comparison) along with a $\sigma_L$ of 0.01 are shown using the dotted lines. The sum of all simulated fluxes equals $10^{-9}$ $MeV^{-1}$ $cm^{-2}$ $s^{-1}$. Flux estimates from the Fermi-LAT 4FGL-DR3 catalog for PWN and MSP sources are also shown using solid lines. Note that the plot above depicts the results of one simulation and shows a simulation different from Figures 2, 3, and 5.

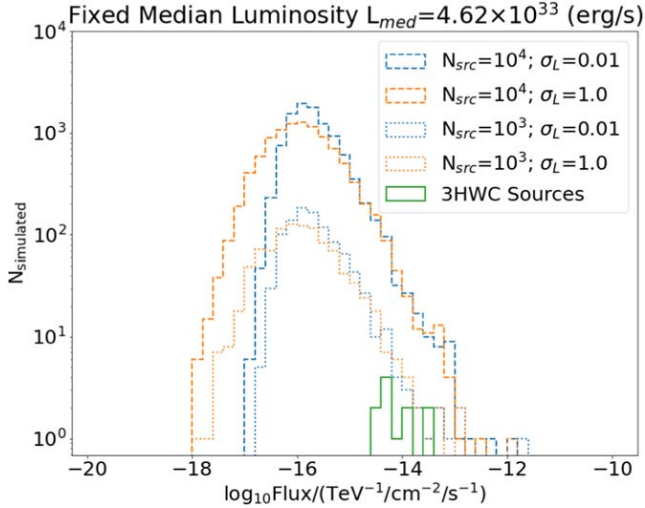

**Figure 5.** Sources simulated using the observations reported in the 3HWC catalog are shown for a $\sigma_L$ of 0.01 and 1.0. The derived differential flux simulations, at an energy of 7 TeV, are shown by dashed lines, while the 11 3HWC source observations (with TeV halo candidate pulsars within 1°) are shown as solid lines. The integrated luminosity over 0.1–100 TeV derived using the 3HWC observations is used as an input in this case, so the total flux estimate is not fixed. As the integrated luminosities derived from the 3HWC catalog are spread out, a higher value of $\sigma_L$ is preferred where the actual observations move toward the brighter end of the simulated histogram. Note that the plot above depicts the results of one simulation and shows a simulation different from Figures 2, 3, and 4.

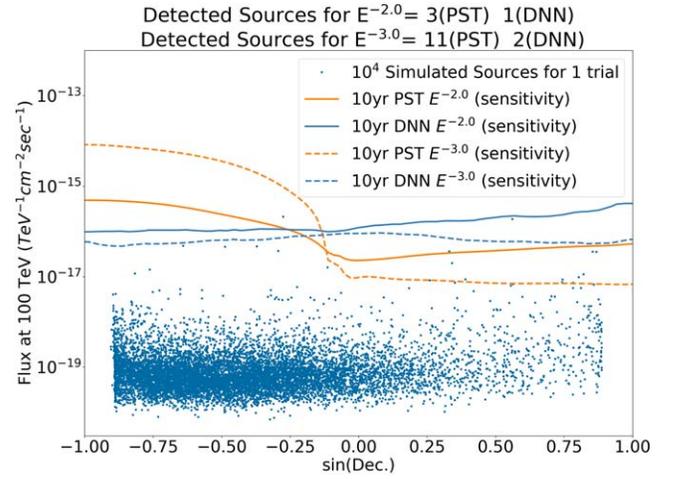

**Figure 6.** Comparison of simulated sources of one sky realization with the 90% CL sensitivity curves from the IceCube PS tracks (Aartsen et al. 2020b) and DNN cascade (Abbasi et al. 2023a) data samples. Blue points show $10^4$ simulated sources with fluxes derived using a LN luminosity distribution with $\sigma_L = 0.01$ (the distribution mimics a SC approximation). If a simulated source flux is above the sensitivity curve, the source is counted as detected. The simulated positions of the sources used in this example simulation are similar to the ones shown in Figure 1.

The number of expected neutrino events also changes as a function of energy because of the power-law index of the neutrino spectrum and the dependence of effective area on neutrino energy. So the total neutrino event counts, integrated over energy, are used, which are then Poisson fluctuated to account for Eddington bias effects. The fluctuated events are compared with a threshold number of events required for detection, which is derived using the sensitivity or discovery potential flux. The number of detected sources for this setup varies as compared to a simple no-bias setup, as shown in Figure 7. More sources for the low-luminosity scenario are detected after the addition of Eddington bias, bringing the simulation closer to the real-world scenario. To provide an estimate for the comparison, if the number of events at $-28°$ required for a detection for an $E^{-2.0}$ spectrum is ~14, then the number of events required for detection reduces to ~8 for an $E^{-3.0}$ spectrum because of the better sensitivity for a $E^{-3.0}$ spectrum using DNN (see Figure 6).

### 5.2. Toy Monte Carlo Setup

Using the SNuGGY framework, we simulate multiple source populations with varying numbers of simulated sources and compare them to IceCube data sets. As the input total diffuse flux is kept constant for these cases, the few source sample is made up of highly luminous sources, while the larger source samples are made up of lower-luminosity sources. We also report the mean luminosity as a function of the number of simulated sources in Figures 8–10, which can be used to put limits on the source population. Using the above method to simulate the sources and after accounting for Eddington bias (see

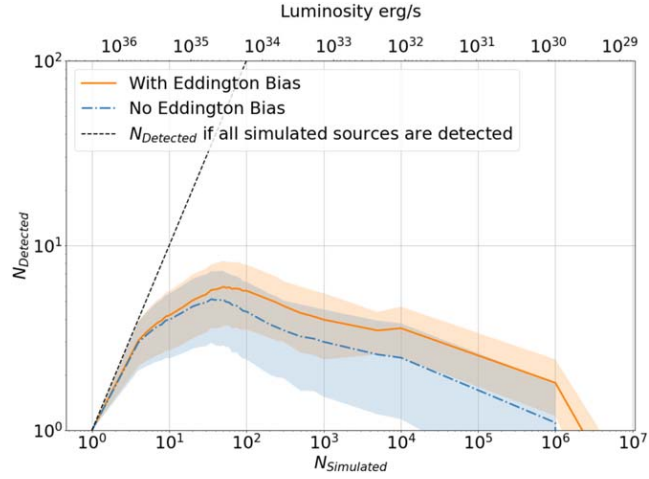

**Figure 7.** Change in the number of detected sources after inclusion of Eddington bias is shown here. Note that the number of detected sources increases in the low-luminosity and high-source-number regime because of fluctuations in the number of detected events due to Eddington bias. We simulate the sources using a LN luminosity function with $\sigma_L = -0.01$ for this check and compare it to the PS tracks data set.

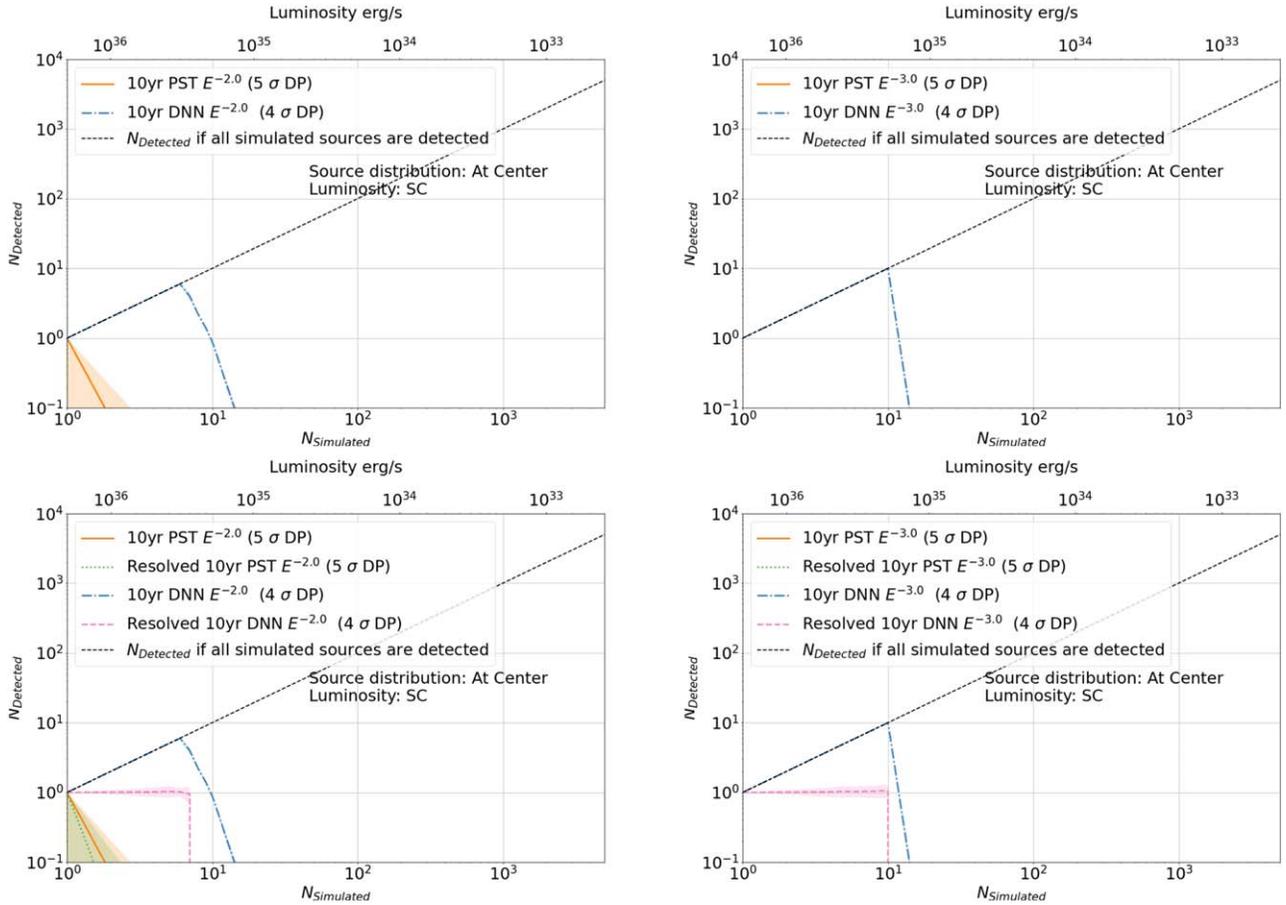

**Figure 8.** Special case for sources simulated at the Galactic center: the number of detected neutrino sources at the Galactic center for different discovery potential curves using $\sigma_L = 0.01$ TeV$^{-1}$ cm$^{-2}$ s$^{-1}$. On the left is for index = 2.0, and right is for index = 3.0. The shaded regions show the $\pm 1\sigma$ uncertainty. Bottom: the number of sources that will be resolved by each data set is also shown along with the uncertainties.

Section 5.1), the number of sources detected is calculated for each case. This is repeated multiple times (at least 1000) to give a mean number of detected sources along with a $1\sigma$ uncertainty. For the cases being tested, we use the sensitivity curves from both the PS tracks samples and the $5\sigma$ ($4\sigma$) discovery potential curves from the PS tracks (DNN) samples. We test the scenarios with the neutrino flux ($dN/dE$) proportional to a power law, with spectral indices of 2.0 and 3.0.

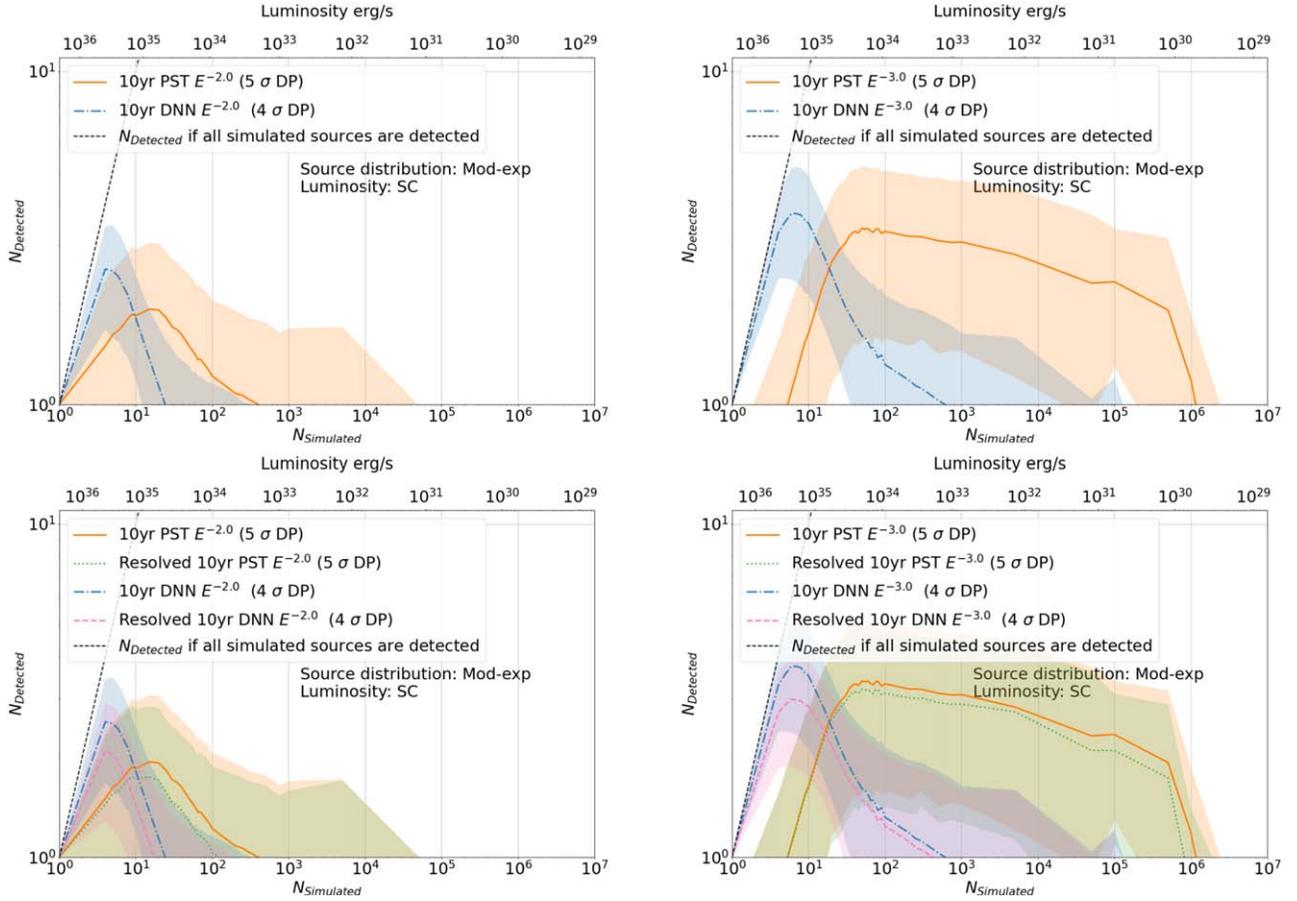

**Figure 9.** Case for sources simulated with a realistic geometric distribution (modified exponential distribution dubbed as "Mod-exp" here) and a SC approach for fluxes: number of detected neutrino sources for different discovery potential curves using $\sigma_L = 0.01$ and the total diffuse flux equals $2.18 \times 10^{-15}$ TeV$^{-1}$ cm$^{-2}$ s$^{-1}$. The $2.18 \times 10^{-15}$ TeV$^{-1}$ cm$^{-2}$ s$^{-1}$ value is obtained using the best-fit neutrino flux derived for the DNN cascade sample using the $\pi^0$ template (Abbasi et al. 2023a). On the left is $E^{-2.0}$ and on the right is $E^{-3.0}$. The shaded regions show the 1$\sigma$ uncertainties. Bottom: the number of sources that will be resolved by each data set is also shown along with the uncertainties.

### 5.3. Case 1: All Sources Are at the Galactic Center

As a test, we simulated sources at the Galactic center and used them to see how the IceCube sensitivity curves compare with them. To achieve this, the simple exponential approach described in Equation (1) is used with an $R_0$ value equal to ∼0.01. Only the radial distance PDF is restricted in this case. The simulated sources are still distributed uniformly around the Galactic center and have different vertical heights ($z$). We then simulate the per-source fluxes by using the LN luminosity model with $\sigma_L = 0.01$ using the method described in Section 3.2. The mean luminosity, in this case, is not specified, allowing the code to estimate it using the method described by Equation (4). Note that this ensures that the mean luminosity changes depending on the number of sources simulated, allowing us to test populations with a few high-luminosity sources versus a large number of low-luminosity sources. The flux distribution is computed once the mean luminosity is estimated based on the value of a given $\sigma_L$.

The results of this test are shown in Figure 8 (discovery potential) and Figure 12 in the Appendix (sensitivity). As the sources are simulated close to the center of the Galaxy along with the SC assumption that they all lie either above or below the sensitivity and discovery potential lines (see Section 6 for more discussion). Note that the results seen for this simulation case come close to the ones shown in Table 1. This confirms the validity of the simulation method used here.

### 5.4. Case 2: Sources Simulated in a More Realistic Scenario

For this case, a more realistic distribution (the "modified exponential," described in Section 3.1) is used to simulate neutrino source locations The fixed parameters, $\alpha = 1.93$, $\beta = 5.06$, and $h = 0.181$ kpc, are used as reported by Ahlers et al. (2016) for the Lorimer et al. (2006) PWN distribution. We also test the parameters for the Case & Bhattacharya (1998) supernova remnant distribution ($\alpha = 2$, $\beta = 3.53$, and $h = 0.181$ kpc), but the final results are very similar to the pulsar distribution case, so we only report the PWN distribution here.

The flux estimation follows a procedure similar to Section 5.3, but with a $\sigma_L$ value equal to 0.01 (SC distribution) and $\sigma_L = 0.5$ (test case to include broader luminosity functions). The results of the $\sigma_L$-0.01 case is shown in Figure 9 (discovery potential) and Figure 13 in the Appendix (sensitivity) while $\sigma_L = 0.5$ case is shown in Figure 10 (discovery potential) and Figure 14 in the Appendix (sensitivity). We use $\sigma_L = 0.5$ (chosen to be close to the value of 0.6 reported by Hooper & Linden 2016) to test the dependence of larger luminosity variance. (See Figure 2 regarding how the simulated luminosities and fluxes change with a change in $\sigma_L$.) The results for this case in terms of the number of sources detected by IceCube are shown in Figures 9–10 (using the discovery potential) and Figures 13–14 in the Appendix (using the sensitivity). Note that the number of sources detected heavily depends on the shape of the sensitivity/discovery potential curve used for comparison. While the sensitivity of the DNN cascade sample is almost similar throughout the whole sky, the sensitivity of the PS tracks data set in the Southern Hemisphere is worse as compared to the Northern Hemisphere. This change in sensitivity becomes even more prominent as the index goes from $-2.0$ to $-3.0$ (see Aartsen et al. 2020b, for more details). For fewer than 10 sources, almost all of the sources are detected using the DNN cascades event sample.

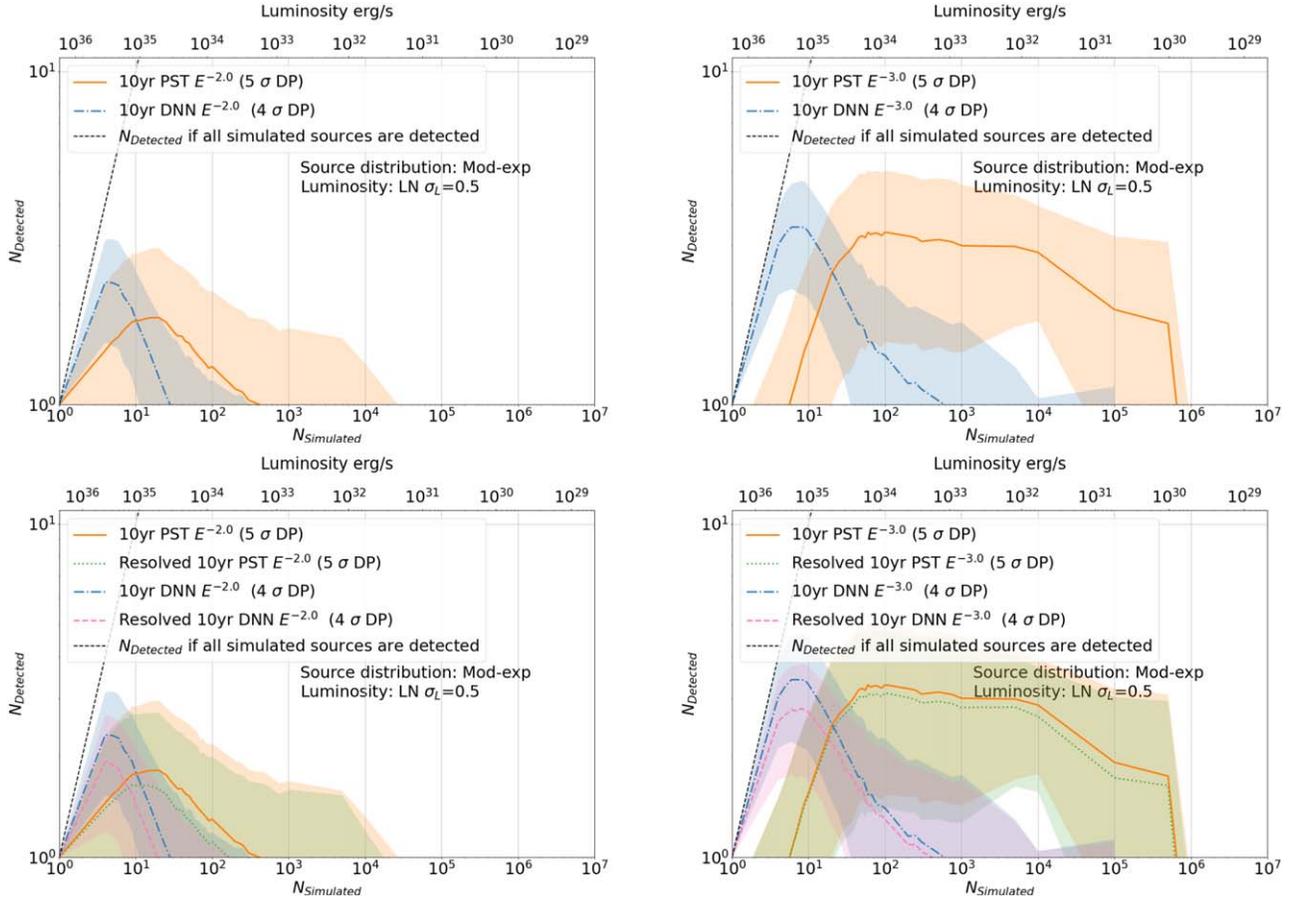

**Figure 10.** Top: case for sources simulated with a realistic geometric distribution (modified exponential distribution dubbed as "Mod-exp" here) and a LN approach for fluxes: number of detected neutrino sources for different discovery potential curves using $\sigma_L = 0.5$ and the total flux equals $2.18 \times 10^{-15}$ TeV$^{-1}$ cm$^{-2}$ s$^{-1}$. On the left is $E^{-2.0}$ and on the right is $E^{-3.0}$. The shaded regions show the $1\sigma$ uncertainties. Bottom: the number of sources that will be resolved by each data set is also shown along with the uncertainties.

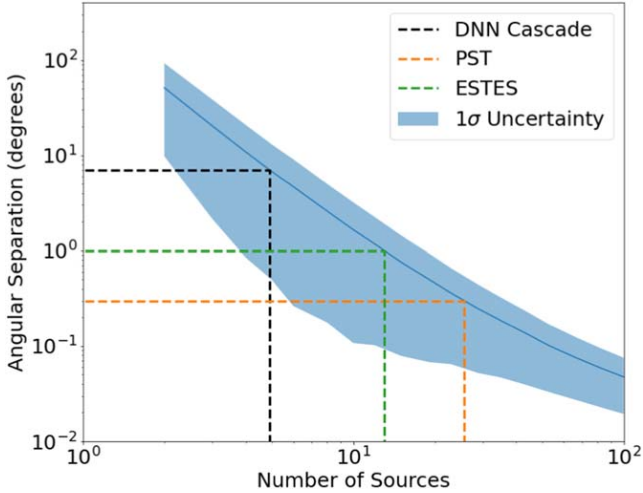

**Figure 11.** Comparison of the minimum separation angle of different number of simulated sources with the angular resolution of IceCube event samples at 100 TeV. The sources are simulated for a modified exponential spatial distribution, as described in Section 5.4. The shaded region shows the $1\sigma$ uncertainties, and the dashed and dotted lines show the minimum number of sources the IceCube event samples will be able to resolve.

### 5.5. Angular Resolution (Source Confusion)

Note that the simulations described above do not account for the angular resolution of the event samples and only check to see if a point source is bright enough to be significant in the IceCube event sample used. This would mean that source confusion effects are not included, and sources that are significant in both the tracks and cascade samples will be noted as "detected" regardless of their separation from one another. While the DNN cascades sample has better sensitivity in the Southern Hemisphere as compared to PS tracks, the angular resolution is poor as compared to the tracks sample (see Abbasi et al. 2023a, for more details).

To test the dependence of our results on angular resolution, we simulate a set of $N$ number of sources similar to Section 5.4 and find the minimum angular separation between the

**Table 2**
Lower Limits on the Approximate Number of Sources Detected by the DNN Cascades Sample Based on the Simulations Are Shown Here (Mean Value of $N_{\text{Detected}}$ in Figures 8–10 Marked by a Blue Dashed Line)

| Quantity | SC ($\sigma_L = 0.01$) | LN ($\sigma_L = 0.5$) | Sources at the center |
|---|---|---|---|
| Not including resolution and source confusion: | | | |
| $N_{\text{src}}(\gamma = 2.0)$ | 25 | 30 | 9 |
| $L_{\text{mean}}(\gamma = 2.0)$ | $3.9 \times 10^{34}$ | $3.2 \times 10^{34}$ | $2.0 \times 10^{35}$ |
| $N_{\text{src}}(\gamma = 3.0)$ | 710 | 683 | 14 |
| $L_{\text{mean}}(\gamma = 3.0)$ | $1.1 \times 10^{33}$ | $1.1 \times 10^{33}$ | $1.5 \times 10^{35}$ |
| After accounting for resolution and source confusion: | | | |
| $N_{\text{src}}(\gamma = 2.0)$ | 19 | 20 | 6 |
| $L_{\text{mean}}(\gamma = 2.0)$ | $6.9 \times 10^{34}$ | $5.0 \times 10^{34}$ | $3.2 \times 10^{35}$ |
| $N_{\text{src}}(\gamma = 3.0)$ | 491 | 416 | 10 |
| $L_{\text{mean}}(\gamma = 3.0)$ | $1.6 \times 10^{33}$ | $1.9 \times 10^{33}$ | $2.0 \times 10^{35}$ |

**Note.** The upper limit on the mean luminosities is also given. This is found by noting the maximum number of simulated sources (dashed line showing the mean value in Figures 8–10 for the discovery potential case) required to detect or resolve at least one source. The top plots in the figures are used to find the values for the detected case (first four rows), and the bottom plots are used to find the detected and resolved values.

simulated sources. This is done ∼100 times to get the mean minimum separation between sources and associated standard deviation. We compare this to the reported angular uncertainty estimates at 100 TeV for DNN cascades (given by ∼7° for all events; see Abbasi et al. 2023a) and PS tracks (given by ∼0.3°; see Abbasi et al. 2021a). The ∼7° resolution is taken using Figure S5 of Abbasi et al. (2023a) with the assumption that the flux is dominated by events with higher energies (<10 TeV). These values are then used to find the number of sources the IceCube event samples will be able to resolve based on the mean simulated minimum separation angle curve. This is shown in Figure 11 where the reported angular uncertainty estimates for different IceCube data sets are compared with the mean simulated minimum separation curve to derive the number of sources the IceCube data set can resolve. We find that DNN cascades will be able to resolve $N \sim 5$ sources using the all events sample, and PS tracks will be able to resolve $N \sim 26$ sources.

Another IceCube data set is currently under development with ∼10 yr of starting tracks events as reported by Silva & Mancina (2019) and Abbasi et al. (2021b). As can be seen from Abbasi et al. (2021b), the shape of the sensitivity curve of this enhanced starting tracks event sample (ESTES) is similar to the cascades sample but with a different scaling. In other words, the sample is equally sensitive in the Northern and Southern Hemispheres. This would imply that the shape of the number of detected ($N_{\text{Detected}}$) curves for DNN cascades shown in Figures 8–10 as blue dashed lines will be similar for ESTES but scaled according to the sensitivity of the ESTES sample. Assuming the angular uncertainty estimates at 100 TeV for ESTES to be 1°.0, we can see from Figure 11 that ESTES will be able to resolve $N \sim 13$ sources.

An additional test is performed to understand better how source confusion can affect the simulation results. To perform this test, we first find the sources that will be detected after accounting for Eddington bias for each simulation. Using the angular uncertainty estimates of the PST and DNN samples, these sets of detected sources are then reduced to remove sources that might be subject to source confusion due to their proximity to other sources. We also consider each source's brightness when calculating this source confusion. This procedure is repeated for each simulation, and the results are shown in Figures 8–10 (bottom row). While the special test case of all the sources at the center is affected the most after accounting for resolution, the results for the other simulations change very slightly, as seen in Figures 8–10 (bottom row). We also see that, for the special case of all sources at the Galactic center, even with source confusion, at least one source will be detected by the DNN cascade sample because of the source's brightness. For the more realistic cases shown in Figures 9 and 10 (bottom), we can see that if the total neutrino background was made up of a few sources (>5 sources), the DNN cascade data set would be able to detect and resolve almost all of the sources.

## 6. Galactic Neutrino Sources Discussion

Using the results shown in Sections 5.3–5.5, we can come to the following conclusions in relation to the Galactic neutrino contributions.

1. *Galactic center sources*. While comparing the Galactic center case (Figure 8) to the other two cases, we can see that because of the special scenario of all the SC sources being simulated close to the center, the uncertainty in the number of detected sources decreases. This is because the uncertainty arises from sources being simulated at different decl., which can put them above or below the IceCube discovery potential curve used for comparison, causing $N_{\text{detected}}$ to change. The DNN cascades sample has a better chance of detecting bright neutrino sources at the Galactic center as compared to PS tracks (see Figure 8). However, in the case of detection, it will be difficult to resolve many sources detected by the cascade sample. The results derived using this test match the ones shown in Table 1 while the simulation rules out the possibility of a small number of very bright sources making up the signal (fewer than 10 sources with luminosities as shown in Figure 8). This is also seen for an index of −3.0, where the DNN sample sensitivity is better than the PS tracks sample sensitivity, meaning that even though PS tracks will be able to detect zero sources, DNN cascades will be able to detect at least ∼8 bright sources (see Table 1 and Figure 8). After accounting for source confusion with this test, we find that DNN can detect and resolve at least one source if the signal is made up of at most seven sources. In the case of no detection using the DNN cascades sample, this scenario of a few bright sources making the background can be ruled out.

2. *Number of neutrinos per source*. The angular separation test shown in Section 5.5 and Figure 11 assumes that sources are resolvable if separated by the median angular resolution of the event sample. Depending on the event selection and source spectrum, detecting a source in a decade or more of integration (and therefore background) requires $O(10)$ neutrino events from the source, which means that the localization of individual detected sources is actually better than the single-event angular resolution and this requirement is conservative. This means that the DNN cascades analysis, for example, would be able to resolve more than eight detected point sources individually if they were from a population following a typical Galactic spatial distribution. This estimate is fortified by the test shown by Figures 9 and 10 (bottom).

3. *Lower limit on the number of Galactic neutrino sources*. The discovery potential comparisons show that if the Milky Way were to have <10 neutrino sources of comparable luminosity ($>10^{35}$ erg s$^{-1}$) producing the total measured flux, the DNN cascades analysis would have detected them. As shown in Table 2, for particular parameter choices, this lower limit on the number of sources is even stronger. We use a luminosity variance of $\sigma_L = 0.5$ to check the dependence of the luminosity function on the number of detected sources. While a significant change is not seen for a spectral index of 2.0, it can be seen for a spectral index of 3.0. The number of detected sources decreases as $\sigma_L$ increases. This is because fewer sources are simulated with high flux values, which will still be detected when compared with the index = 3.0 sensitivity and discovery potential curves (see also Figure 6). This is expected for populations with larger luminosity variance of $\sigma_L = 0.5$ (close to the value of 0.6 reported by Hooper & Linden 2016) as compared to 0.01. As the total number of Galactic sources increases, the mean luminosity decreases, and fewer sources would be detected with the DNN cascades event sample. This holds true even after including fluctuations due to Eddington bias, which increases the number of detected sources in the regime of many low-luminosity sources. To put a conservative limit on the number of detected sources using the DNN cascades sample, we use the SC scenario with a modified exponential distribution (column 2 in Table 2). For an index of 2.0, we find a limit of 25 sources for the detected scenario and 19 sources for the resolved one. After accounting for uncertainties for $N_{\rm Detected}$ (see Figure 9, left panel) to find the point where $N_{\rm Detected}({\rm mean} - {\rm uncertainty}) = 1$, changes this limit to ∼12 sources (detected case) and ∼8 sources (resolved case). Using these results from our simulation study and the fact the Abbasi et al. (2023a) were not able to detect any sources, we can conclude that there are ≳8 sources making up the diffuse signal with a mean luminosity of $1.6 \times 10^{35}$ erg s$^{-1}$ of the sample.

## 7. Conclusion

This work describes the SNuGGY simulation code, which can be used to simulate Galactic point sources along with their neutrino and gamma-ray fluxes. The diversity of the analyses that can be performed using the code is shown, along with a focus on using the simulation to draw robust conclusions from the recent IceCube (Abbasi et al. 2023a) detection of the Galactic neutrino flux. Using a Monte Carlo simulation performed for Galactic neutrino sources, we determine lower limits on the number of Galactic sources contributing to the observed flux. As the distribution of different Galactic source classes follows a similar pattern, the simulations presented here can be applied to different Galactic source classes and even be used as points of neutrino emission in the Galaxy due to cosmic rays interacting with matter. Nondetection of any individual Galactic neutrino source to date (Kheirandish & Wood 2019; Aartsen et al. 2020a; Abbasi et al. 2022b, 2023a, 2023c), combined with the total flux normalization measured by IceCube, enable us to determine that there must be more than ∼8 individual sources (or points of diffuse emission), each with a luminosity of $1.6 \times 10^{35}$ erg s$^{-1}$, producing the signal.

## Acknowledgments

The authors would like to thank the members of the IceCube Collaboration for their valuable suggestions. The authors would also like to thank Markus Ahlers for helpful discussions relating to the spatial distribution of the Galactic sources, Alex Pizzuto for helping with the initial development of the SNuGGY code, and Aswathi Balagopal and Sam Hori for discussions related to the result. During this work, A.D. was supported by the John Bahcall Fellowship at Wisconsin IceCube Particle Astrophysics Center (WIPAC) at the University of Wisconsin-Madison. J.V. is supported by a Vilas Associate award at the University of Wisconsin-Madison.

## Appendix

We make use of discovery potential curves for the study as shown in Table 1 and Figures 8–10. Here, we also show how the above analysis changes when sensitivity estimates for the IceCube data sets are used instead of discovery potential. In Table 3, we report how $N_{\rm src}$ changed when sensitivity curves are used instead of discovery potential. Figures 12–14 show the results using sensitivity curves instead of discovery potential.

**Table 3**
Similar to Table 1 but Using Sensitivities Instead of the Discovery Potential

| Sample Tested | $E^{-2.0}$ Flux at ∼−28° | $>N_{\rm src}$ ($E^{-2.0}$) | $E^{-3.0}$ Flux at ∼−28° | $>N_{\rm src}$ ($E^{-3.0}$) |
|---|---|---|---|---|
| DNN sensitivity | $1.05 \times 10^{-16}$ | 14 | $6.19 \times 10^{-17}$ | 24 |
| PST sensitivity | $2.25 \times 10^{-16}$ | 7 | $3.29 \times 10^{-15}$ | 0 |

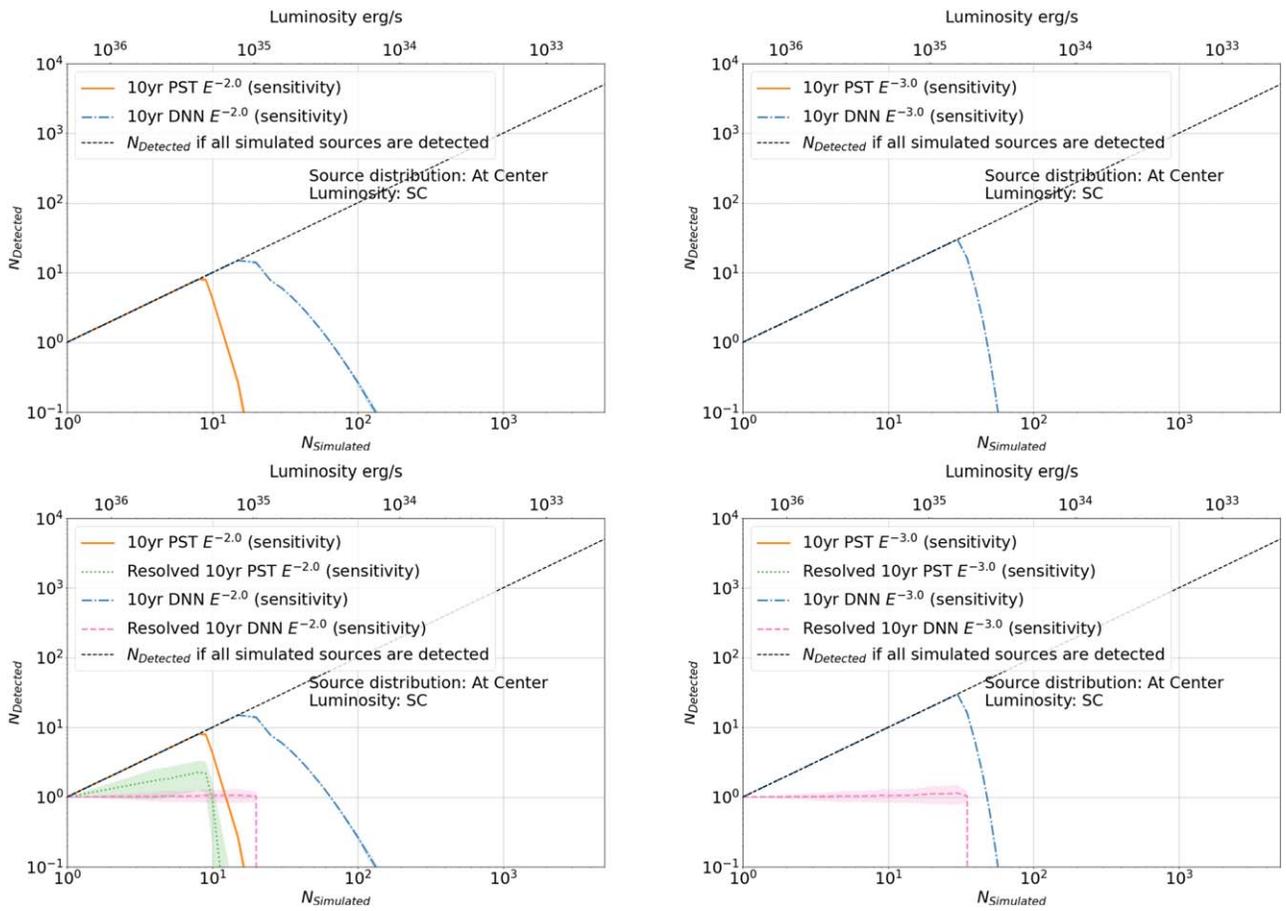

**Figure 12.** Same as Figure 8 but using sensitivity curves.

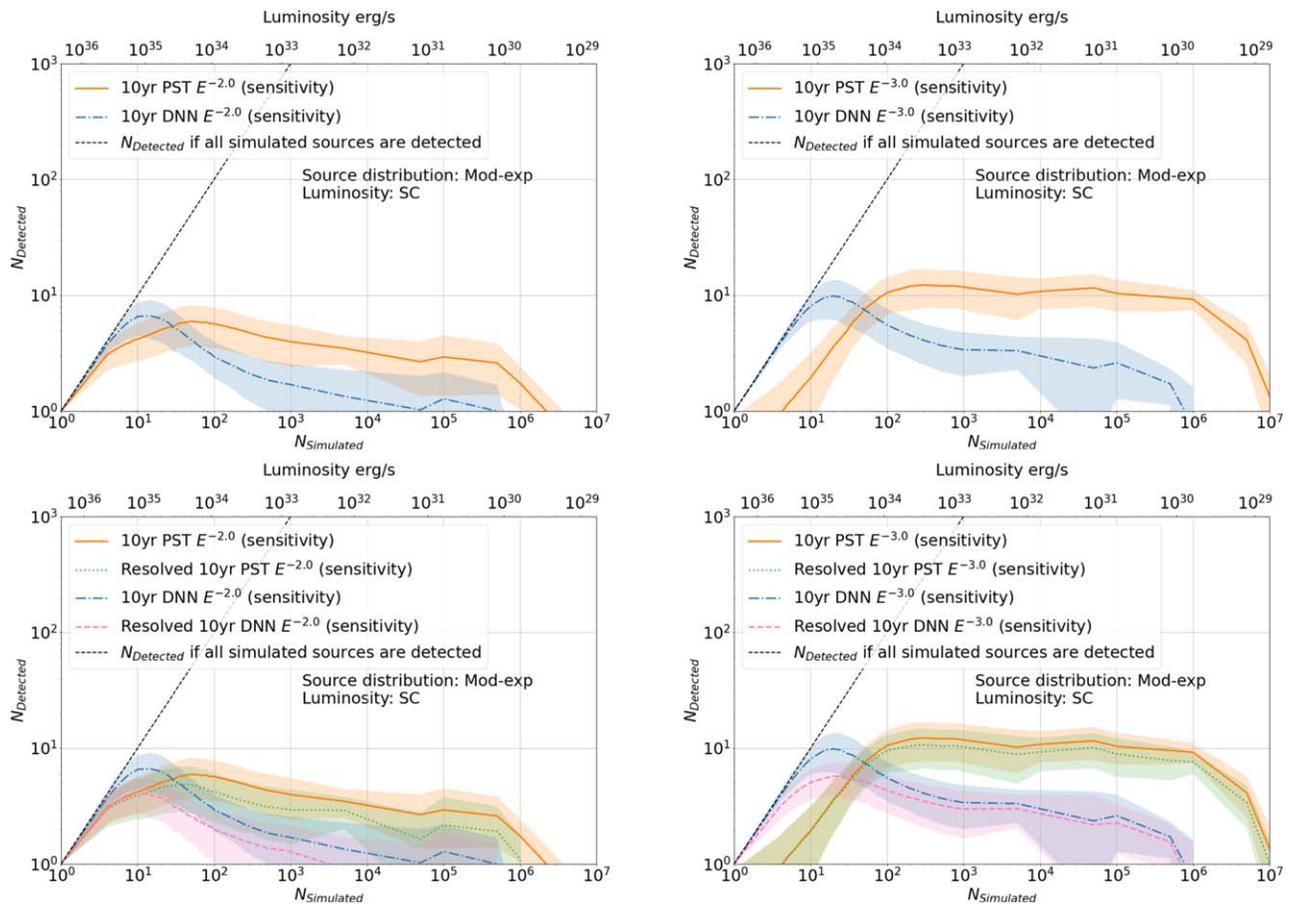

**Figure 13.** Same as Figure 9 but using sensitivity curves.

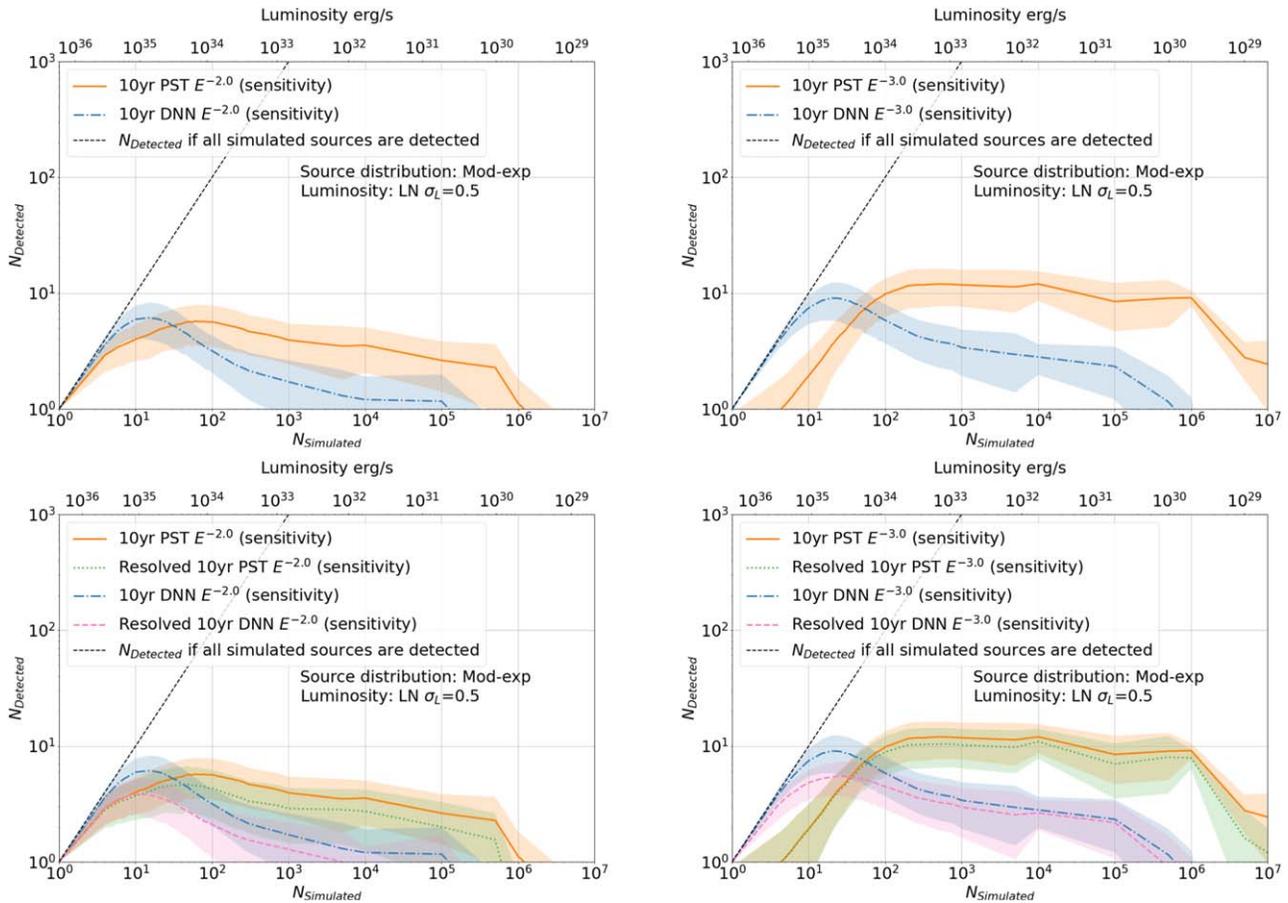

Figure 14. Same as Figure 10 but using sensitivity curves.

## ORCID iDs

Abhishek Desai ● https://orcid.org/0000-0001-7405-9994
Justin Vandenbroucke ● https://orcid.org/0000-0002-9867-6548
Samalka Anandagoda ● https://orcid.org/0000-0002-5490-2689
Jessie Thwaites ● https://orcid.org/0000-0001-9179-3760
M. J. Romfoe ● https://orcid.org/0009-0001-6042-701X